\documentclass[9pt,twocolumn,twoside]{opticajnl}
\journal{opticajournal}

\setboolean{shortarticle}{true}

\usepackage{amsfonts}
\usepackage{bm}

\usepackage{physics}

\newcommand{\Esp}{\mathbb{E}}
\newcommand{\RR}{\mathbb{R}}

\usepackage{lineno}

\title{Robust optical design and closed-form tolerancing through autodiff-based Hessian spectral analysis}

\author[1,*]{Bastien Laville}
\author[1]{Benjamin Aymard}

\affil[1]{Thales Alenia Space, 5 All\'{e}e des Gabians, 06150 Cannes, France}
\affil[*]{bastien.laville@thalesaleniaspace.com}

\begin{abstract}
  Robust optical design demands quantitative knowledge of how
  manufacturing and alignment tolerances degrade system performance.
  We show that
  analysing the perturbation eigenmodes of the Hessian matrix gives
  qualitative insight about the mechanisms of performance degradation
  (such as couplings) that is invisible to classical sensitivity-matrix
  analysis based on the Jacobian alone.
  Via the envelope theorem, we prove that the first-order sensitivity
  of the fully compensated system is identical to that of the
  uncompensated one; refocusing only acts at second order through the
  Schur complement of the Hessian.
  We propose the trace of the tolerance-scaled Hessian
  as a single scalar
  robustness metric.
  Demonstrated on an off-axis three-mirror anastigmat and scaled to a
  twenty-three-parameter surface-figure model, eigenmode decomposition
  reveals the dominant sensitivity directions and yields deterministic
  tolerance budgets that replace costly Monte Carlo sampling.
\end{abstract}

\setboolean{displaycopyright}{false}

\begin{document}

\maketitle

\section{Introduction}

Optical tolerancing strives to predict how manufacturing and alignment
imperfections degrade imaging performance, and traditionally relies on
Monte Carlo (MC) simulation~\cite{bates_influence_2004}. The standard
workflow perturbs each degree of freedom (DoF) according to a
manufacturing distribution, refocuses compensators, and collects the
statistical spot-size degradation over $N$ random draws. However,
this approach suffers from a severe sample-complexity bottleneck.
Non-parametric tolerance intervals based on the Wilks
formula~\cite{wilks_determination_1941} require
$N\!\geq\!\ln(1{-}C)/\ln(p)$ samples to bound the $p$-th percentile
at confidence~$C$; for 99.9\% yield at 99\% confidence,
$N\!\geq\!4{,}602$. Moreover, with $k$ independent perturbations, the
probability of probing the worst-case corner (say, outer 10\% of each axis
simultaneously) is $(0.1)^k$; for a three-mirror anastigmat (TMA) telescope
with $k\!=\!9$ DoFs, MC requires ${\sim}10^9$ samples to explore this
corner at 99\% confidence, hence an intractable proposition.

Differentiable ray
tracing~\cite{sun_end--end_2021,wang_do_2022,tseng_differentiable_2021}
emerges as a powerful alternative: automatic differentiation
(AD)~\cite{baydin_automatic_2018} delivers exact derivatives of
optical metrics at the cost of a single reverse-mode (adjoint) pass,
which propagates sensitivities backward through the ray-trace and yields
the full gradient with respect to all $N$ parameters at a cost
comparable to one extra trace, independent of $N$. Recent work
has exploited these gradients for lens
design~\cite{yang_curriculum_2024} and sensitivity
analysis~\cite{wang_do_2022}, but the second-order Hessian
structure, and its implications for robust design, remains largely
under-exploited, despite recent works in this direction~\cite{hu_economic_2022}.
In this Letter, we derive a theoretical framework
connecting the Hessian of the performance metric to robust design
objectives. We prove via the envelope
theorem~\cite{milgrom_envelope_2002} that first-order sensitivity is
invariant under active refocusing, propose
$\Tr(\mathbf{S}\mathbf{H}\mathbf{S})$ as a physically motivated
scalar robustness measure, and demonstrate our pipeline on off-axis
TMAs using a JAX-based~\cite{jax_2018} differentiable engine.

\section{Theory}
\label{sec:theory}

\subsection{Performance function and Taylor expansion}

Consider an optical system parameterised by a vector of structural
parameters $\bm{p} = (p_1,\ldots,p_N)\in\RR^N$ (such as mirror tilts,
decentres, inter-mirror spacings, and surface curvatures).
The imaging performance is evaluated using a scalar metric
$\mathcal{P}(\bm{p})$

Assuming that the scalar metric is regular,
around a nominal design point $\bm{p}^*$ obtained by optimisation,
the Taylor expansion of the performance metric reads:
\begin{equation}\label{eq:taylor}
  \mathcal{P}(\bm{p}^*\!+\!\delta\bm{p})
  \;\approx\;
  \underbrace{\mathcal{P}(\bm{p}^*)}_{\text{nominal}}
  +
  \underbrace{\bm{G}^\top\!\delta\bm{p}}_{\text{sensitivity}}
  +
  \tfrac{1}{2}\,
  \underbrace{\delta\bm{p}^\top\!\mathbf{H}\,\delta\bm{p}}_{\text{coupling/curvature}},
\end{equation}
where $\bm{G} = \nabla_{\!\bm{p}}\mathcal{P}(\bm{p}^*)$ is the
gradient vector and $\mathbf{H} =
\nabla_{\!\bm{p}}^2\mathcal{P}(\bm{p}^*)$ is the symmetric $N \times
N$ Hessian matrix.

\emph{Remark.}
In this context, curvature refers to the second-order derivative with respect to a
parameter, not to the lens/mirror curvature.

At a local design optimum, the nominal gradient vanishes
$\bm{G}=\bm{0}$; the degradation under small perturbations is then
purely quadratic and entirely characterised by the Hessian $\mathbf{H}$.
Strong misalignment and/or vignetting both make the performance metric
non-smooth, and the method no longer applies.

\subsection{Hessian matrix and perturbation eigenmodes}
Diagonalising the tolerance-scaled Hessian $\mathbf{H}_{\text{sc}} =
\mathbf{S}\mathbf{H}\mathbf{S} =
\mathbf{V}\boldsymbol{\Lambda}_{\text{sc}}\mathbf{V}^\top$ with
eigenvalues
$\boldsymbol{\Lambda}_{\text{sc}}=\mathrm{diag}(\lambda_1^{\text{sc}},\ldots,\lambda_N^{\text{sc}})$
yields orthogonal \emph{perturbation eigenmodes}~$\bm{v}_i$. A
tolerance-scaled perturbation $\delta\tilde{\bm{p}} =
\mathbf{S}^{-1}\delta\bm{p} = \sum_i \alpha_i\bm{v}_i$ decomposes the
performance degradation as:
\begin{equation}\label{eq:eigen}
  \Delta\mathcal{P}
  \;=\;
  \tfrac{1}{2}\sum_{i=1}^{N}\alpha_i^2\lambda_i^{\text{sc}},
\end{equation}
separating the sensitivity of the system into independent deformation channels.

\subsection{Compensator invariance via the envelope theorem}

When a physical perturbation $\delta\bm{p}$ occurs,
active compensators $\bm{c}\in\RR^n$ (such as detector translation
stages, refocusing actuators, or active mirror controllers) are
re-optimised to recover imaging quality.
Let $L(\bm{c},\bm{p})$ denote the loss function with both compensator
settings $\bm{c}$ and manufacturing/alignment perturbations $\bm{p}$
explicit, and let $\bm{c}^*(\bm{p}) = \arg\min_{\bm{c}}
L(\bm{c},\bm{p})$ denote the optimal compensator setting for any
given perturbation $\bm{p}$.
Under active compensation, the re-optimised system performance is
tracked by the compensated loss function $L^*(p) = L(\bm{c}^*(\bm{p}), \bm{p})$.
At any perturbation state in a neighbourhood of the nominal,
$\bm{c}^*(\bm{p})$ satisfies the first-order necessary condition of
stationarity, for all $i = 1, \ldots, n$:

\begin{equation}\label{eq:stationarity}
  \nabla_{\!\bm{c}} L(\bm{c}^*(\bm{p}), \bm{p}) = \bm{0} \quad
  \implies \quad \pdv{L}{c_i}(\bm{c}^*(\bm{p}), \bm{p}) = 0.
\end{equation}

To evaluate the first-order sensitivity of the fully compensated
system to perturbations, we differentiate $L^*(\bm{p})$ using the
multivariate chain rule:

\begin{equation}\label{eq:compensated_gradient}
  \dv{L^*}{p_k} = \pdv{L}{p_k} + \sum_{i=1}^n
  \underbrace{\pdv{L}{c_i}}_{=0} \pdv{c_i^*}{p_k} = \pdv{L}{p_k},
\end{equation}

which in vector form reads $\dv{L^*}{\bm{p}} \eval_{\bm{p}^*} =
\nabla_{\!\bm{p}} L$.

Hence, \emph{at first order, the sensitivity of the fully refocused system
is identical to that of the uncompensated system}: the uncompensated
gradient from a single backward pass is the exact compensated Jacobian,
and refocusing acts strictly on curvature.
Differentiating once more (full derivation in Supplement, Sec.~S2)
shows that the compensated second-order behaviour is the Schur
complement of the compensator block,
\begin{equation}\label{eq:schur}
  \mathbf{H}_{\text{comp}} = \mathbf{H}_{pp} -
  \mathbf{H}_{pc}\,\mathbf{H}_{cc}^{-1}\,\mathbf{H}_{cp},
\end{equation}
where $\mathbf{H}_{pp}$, $\mathbf{H}_{pc}$, and $\mathbf{H}_{cc}$ are
blocks of the full system Hessian over $(\bm{p}, \bm{c})$. Physically,
$\mathbf{H}_{cc}$ is the \emph{refocusing stiffness} (how sharply the
loss rises as a compensator moves off its optimum), $\mathbf{H}_{pc}$
is the \emph{cross-coupling} between perturbations and compensators, and
$\mathbf{H}_{pp}$ is the raw, uncompensated curvature. The subtracted
term $\mathbf{H}_{pc}\mathbf{H}_{cc}^{-1}\mathbf{H}_{cp}$ is therefore
the sensitivity \emph{bought back} by the stage: a perturbation that
couples strongly into a cheap-to-move compensator (small
$\mathbf{H}_{cc}$) has most of its curvature removed. The effective
compensated sensitivity is then $\Delta\mathcal{P} \approx
\bm{G}^\top \delta\bm{p} +
\tfrac{1}{2}\delta\bm{p}^\top \mathbf{H}_{\text{comp}} \delta\bm{p}$.

\subsection{Hessian-based tolerancing}

For independent manufacturing perturbations
$p_i\sim\mathcal{U}(-w_i,w_i)$, let $\mathbf{S} =
\mathrm{diag}(\sigma_1,\ldots,\sigma_N)$ with $\sigma_i =
w_i/\sqrt{3}$ (standard deviation of the uniform distribution).
The expected performance degradation under the quadratic model is
\begin{equation}\label{eq:expected}
  \Esp[\Delta\mathcal{P}]
  \;=\;
  \bm{G}^\top\!\Esp[\delta\bm{p}]
  \;+\;
  \tfrac{1}{2}\Tr\bigl(\mathbf{H}\,\mathrm{Cov}[\delta\bm{p}]\bigr)
  \;=\;
  \tfrac{1}{2}\Tr(\mathbf{S}\mathbf{H}\mathbf{S}),
\end{equation}
since $\Esp[\delta\bm{p}]=\bm{0}$ and $\mathrm{Cov}[\delta\bm{p}] =
\mathbf{S}^2$ for independent variables.
This identity depends only on the second moment of the perturbations:
for any zero-mean distribution
\[
  \Esp[\Delta\mathcal{P}] =
  \tfrac{1}{2}\Tr(\mathbf{H}\,\mathrm{Cov}[\delta\bm{p}]),
\]
so correlated tolerances are accommodated by replacing $\mathbf{S}^2$ with the full
covariance $\boldsymbol{\Sigma}$ (i.e.\ $\mathbf{S}\!\to\!\boldsymbol{\Sigma}^{1/2}$),
while the distributional shape (e.g.\ Gaussian) affects only the
higher-order tail, not the mean.

\subsection{Robust design objective}

This motivates the \emph{robust design objective}, for $\gamma > 0$:
\begin{equation}\label{eq:objective}
  \boxed{J
    \;=\;
    \mathcal{P}(\bm{p})
    \;+\;
  \gamma\,\Tr(\mathbf{S}\mathbf{H}\mathbf{S}),}
\end{equation}
which minimises both the nominal performance and the expected
quadratic degradation under tolerances.
The trace formulation has three practical advantages over the
eigenvalue-based $\sum\lambda_i^2$:
(i)~$\Tr(\mathbf{S}\mathbf{H}\mathbf{S})=\sum_i \sigma_i^2 H_{ii}$ requires only
diagonal Hessian entries, computable by $N$ forward-over-reverse AD
passes without full eigendecomposition. For ultra-high-dimensional cases
(e.g., freeform surfaces where $N \gg 10^3$), this $N$-pass diagonal cost
can be bypassed using Hutchinson's randomized trace
estimator~\cite{hutchinson_stochastic_1989}
combined with JAX's vectorized mapping (\texttt{vmap}), reducing complexity
to $M \ll N$ passes with random probe vectors;
(ii)~it is numerically stable even when eigenvalues span many orders
of magnitude;
(iii)~it avoids third-order AD (needed for $\nabla\sum\lambda_i^2$).


\subsection{Statistical mechanics analog}

Equation~\eqref{eq:eigen} is formally the \emph{equipartition} of a
quadratic energy: each perturbation eigenmode contributes
$\tfrac{1}{2}\alpha_i^2\lambda_i^{\text{sc}}$, in analogy with the
$\tfrac{1}{2}k_B T$ carried by every quadratic degree of freedom in
thermal equilibrium (developed in Supplement, Sec.~S5). The design
consequence is what matters here: a nominal optimum typically has a
highly uneven sensitivity spectrum dominated by a few exceptionally
sensitive ``weak directions'', and minimising
$\Tr(\mathbf{S}\mathbf{H}\mathbf{S})$ acts as a regulariser that flattens
this spectrum toward an equipartition-like state, wherever the design
admits robustness slack. Conversely, when the dominant sensitivity is
locked to the prescription (e.g.\ system focal length) it is
structurally irreducible and must be met by manufacturing precision
instead (cf.\ Sec.~\ref{sec:demo}E). When some modes are correctable
during assembly, the objective should act on the Schur complement
(Supplement, Sec.~S2) of the uncompensated DoFs in~\eqref{eq:schur}, not
the full Hessian.

\section{Numerical demonstration}
\label{sec:demo}

\subsection{TMA system and analytical Hessian results}

We consider an off-axis TMA
telescope~\cite{cook_three-mirror_1979,korsch_anastigmatic_1977} with
five geometry DoFs: three mirror tilts $(\theta_1,\theta_2,\theta_3)$
and two inter-mirror spacings $(d_1,d_2)$.
The performance metric $\mathcal{P}$ is the centroid-subtracted
mean-squared spot size, averaged over 9 field points with 200 rays per field.

Exact analytical derivatives are obtained by attaching a custom
Jacobian--vector product to each ray--surface intersection via the
implicit function theorem, evaluated once at the converged root, rather
than by unrolling (back-propagating through) the iterative Newton
solver which is both costly and numerically unstable. This requires
the solver to be converged far more tightly than for spot diagrams;
once it is, the AD derivatives agree with central finite differences to
better than $10^{-9}$ relative and the Hessian is symmetric to
$<10^{-14}$ (see Supplement, Sec.~S1 and~S4).

The Hessian computation requires one call to
\texttt{jax.hessian(\,$\mathcal{P}$\,)} (forward-over-reverse AD).
The resulting Hessian $\mathbf{H} = \nabla^2\mathcal{P}(\bm{p}^*)$
has symmetry error $\|\mathbf{H}-\mathbf{H}^\top\|_\infty /
\|\mathbf{H}\|_\infty < 10^{-14}$, four positive eigenvalues,
and one negligibly negative eigenvalue ($-3.94 \times 10^{-13}$~m$^2$)
representing a numerically flat direction, confirming that $\bm{p}^*$
is effectively at a local minimum.

\begin{table}[t]
  \centering
  \footnotesize
  \setlength{\tabcolsep}{3pt}
  \caption{Tolerance-scaled eigenvalues $\lambda_i^{\text{sc}}$ and
    dominant eigenvector components of $\mathbf{H}_{\text{sc}} =
    \mathbf{S}\mathbf{H}\mathbf{S}$.
    The expected spot-size change per unit-tolerance perturbation
    along each mode is $\Delta\mathcal{P}_{1\sigma} =
    \tfrac{1}{2}|\lambda_i^{\text{sc}}|$.
    Eigenvector components are dimensionless: they are taken in the
    tolerance-normalised (unit-variance) coordinate
    $\tilde{\bm{p}}=\mathbf{S}^{-1}\bm{p}$, so tilt and spacing DoFs are
    directly comparable.
  }
  \label{tab:eigenmodes}
  \begin{tabular}{cccl}
    \hline
    Mode & $\lambda_i^{\text{sc}}$ [m$^2$] &
    $\Delta\mathcal{P}_{1\sigma}$ [$\mu$m$^2$] & Dominant DoFs \\
    \hline
    1 & $+4.98 \times 10^{-10}$ & $249.23$ & $d_1$ ($-0.81$), $\theta_1$ ($-0.44$),
    $\theta_2$ ($-0.34$) \\
    2 & $+2.69 \times 10^{-11}$ & $13.46$  & $\theta_2$ ($+0.66$), $d_1$ ($-0.56$),
    $\theta_1$ ($+0.48$) \\
    3 & $+7.68 \times 10^{-13}$ & $0.38$   & $\theta_2$ ($+0.58$), $\theta_1$ ($-0.58$),
    $d_2$ ($+0.41$) \\
    4 & $+3.71 \times 10^{-14}$ & $0.02$   & $\theta_3$ ($+0.71$), $d_2$ ($+0.68$),
    $\theta_1$ ($-0.13$) \\
    5 & $-2.25 \times 10^{-15}$ & $0.00$   & $d_2$ ($+0.60$), $\theta_3$ ($-0.55$),
    $\theta_1$ ($+0.47$) \\
    \hline
  \end{tabular}
\end{table}

Table~\ref{tab:eigenmodes} presents the perturbation eigenmodes.
The condition number
$|\lambda_1^{\text{sc}}|/|\lambda_5^{\text{sc}}| = 2.2 \times 10^5$
reveals strongly anisotropic sensitivity: the dominant mode couples
$d_1$ with $\theta_1, \theta_2$ pointing to M1–M2 focal-length co-dependence,
while the weakest mode is a near-flat direction dominated by $d_2$.
The scalar robustness measure is $\Tr(\mathbf{S}\mathbf{H}\mathbf{S})
= 5.26 \times 10^{-10}$~m$^2$ ($526.19$~$\mu$m$^2$).

\subsection{Taylor-expansion validation}

\begin{figure}[t]
  \centering
  \includegraphics[width=0.68\linewidth]{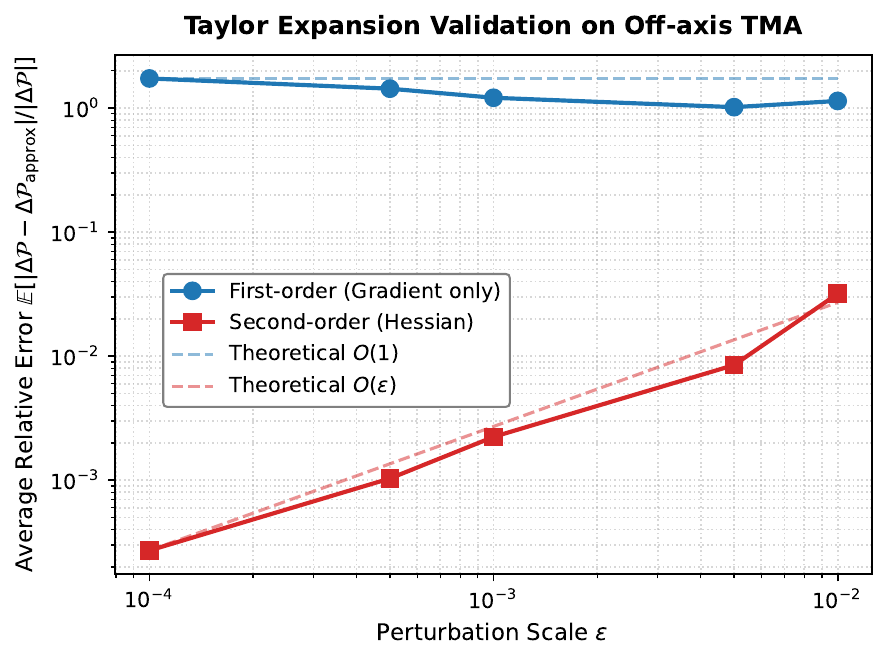}%
  \vspace{-0.25cm}
  \caption{Relative error of the Taylor expansion~\eqref{eq:taylor}
    vs perturbation scale~$\varepsilon$.
    Blue circles: first-order (gradient only); red squares:
    second-order (gradient + Hessian).
    Dashed lines: theoretical $\mathcal{O}(1)$ and $\mathcal{O}(\varepsilon)$ slopes.
  Each point averages over 1,000 random directions.}
  \vspace{-0.35cm}
  \label{fig:taylor}
\end{figure}

To validate the quadratic model, we evaluate
$\mathcal{P}(\bm{p}^*\!+\!\varepsilon\,\hat{\bm{v}})$ for random unit
directions $\hat{\bm{v}}$ and eigenmodes, at perturbation scales
$\varepsilon\in[10^{-4},\!10^{-2}]$.
Figure~\ref{fig:taylor} shows the relative approximation error.
Because $\bm{p}^*$ is a local minimum, $\nabla\mathcal{P}(\bm{p}^*){\approx}0$
and the gradient-only approximation carries $\mathcal{O}(1)$ relative error:
the linear correction is negligible and $\Delta\mathcal{P}$ is dominated by
the quadratic term.
Including the Hessian reduces the residual to $\mathcal{O}(\varepsilon)$,
confirming the envelope theorem insight and that the quadratic model captures the
dominant sensitivity over the
relevant tolerance range.

\subsection{Closed-form yield versus Monte Carlo}

\begin{figure}[t]
  \centering
  \includegraphics[width=0.68\linewidth]{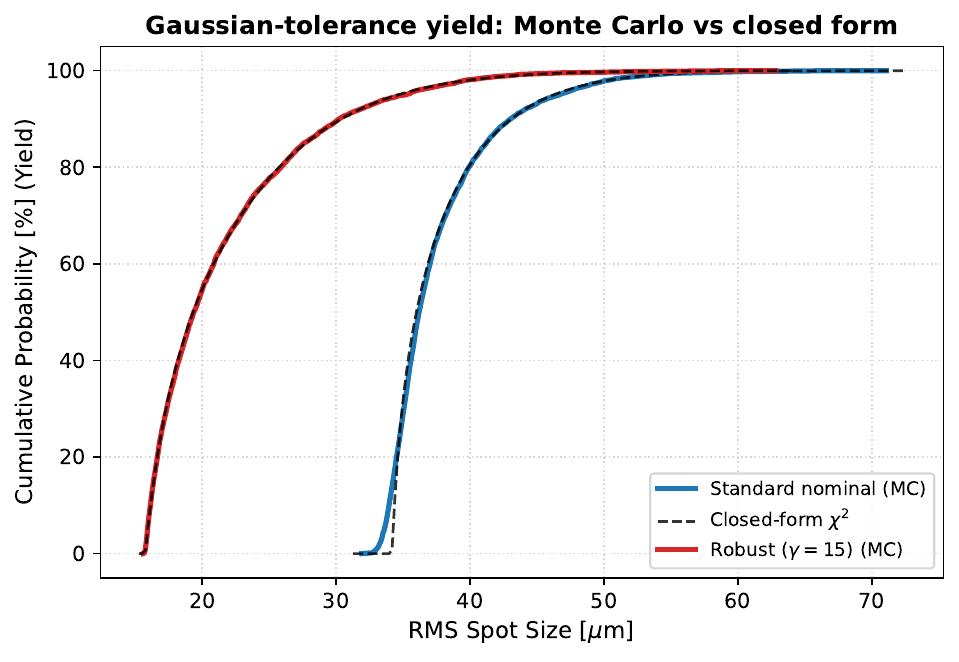}%
  \vspace{-0.25cm}
  \caption{Gaussian-tolerance yield (CDF of RMS spot size) for the
    standard nominal (blue) and robust ($\gamma=15$, red) designs from a
    5,000-sample ray-traced Monte Carlo (solid), with the closed-form
    generalised-$\chi^2$ prediction, computed from the perturbation
    eigenvalues alone, no ray tracing, overlaid (black dashed) on each.
    The robust design improves the $99.9\%$ threshold from $63.11$ to
  $55.70\;\mu$m.}
  \vspace{-0.35cm}
  \label{fig:yield}
\end{figure}

The key advantage of the Hessian model is that, for Gaussian
tolerances, the entire yield curve is available in \emph{closed form}.
At a local optimum ($\bm{G}\!\approx\!\bm{0}$), writing the tolerances
as $\delta\bm{p}=\mathbf{S}\bm{z}$ with $\bm{z}\!\sim\!\mathcal{N}(\bm 0,\mathbf{I})$
and diagonalising $\mathbf{H}_{\text{sc}}=\mathbf{V}\boldsymbol{\Lambda}_{\text{sc}}\mathbf{V}^\top$
turns the degradation into a \emph{generalised chi-squared} $\chi^2$,
\begin{equation}\label{eq:genchi2}
  \Delta\mathcal{P} \;=\; \tfrac{1}{2}\sum_{i=1}^{N}
  \lambda_i^{\text{sc}}\,w_i^2, \qquad w_i\!\sim\!\mathcal{N}(0,1)\ \text{i.i.d.},
\end{equation}
a weighted sum of independent $\chi^2_1$ variables whose weights are
exactly the perturbation eigenvalues $\lambda_i^{\text{sc}}$. Its CDF
has no elementary form for distinct weights, but is recovered exactly by
a one-dimensional inversion of the characteristic function
$\varphi(u)=\prod_i(1-\mathrm{i}\lambda_i^{\text{sc}}u)^{-1/2}$
\cite{imhof_computing_1961} broken down into its polar form:
\begin{equation}\label{eq:imhof}
  \Pr(\Delta\mathcal{P} > x) = \tfrac{1}{2} + \frac{1}{\pi}\!\int_0^\infty
  \mathrm{d}u \, \frac{\sin\theta(u)}{u\,\rho(u)},
\end{equation}
with $\theta(u)=\tfrac{1}{2}\sum_i\arctan(\lambda_i^{\text{sc}}u)-\tfrac{1}{2}xu$
and $\rho(u)=\prod_i[1+(\lambda_i^{\text{sc}})^2u^2]^{1/4}$ (derivation
in Supplement, Sec.~S3). Equation~\eqref{eq:imhof} needs only the $N$
eigenvalues (no ray tracing) and a single quadrature; the yield curve
of Fig.~\ref{fig:yield} is its percentiles mapped through
$\mathrm{RMS}=\sqrt{\mathcal{P}(\bm{p}^*)+\Delta\mathcal{P}}$. For both
designs it reproduces a $5{,}000$-sample ray-traced Gaussian Monte Carlo
to within $1.4\;\mu$m at every percentile, while being orders of
magnitude faster and free of sampling noise.

Two scope conditions matter. First, while the mean degradation
\eqref{eq:expected} holds for \emph{any} zero-mean distribution, the
closed-form CDF \eqref{eq:imhof} relies specifically on the
perturbations being Gaussian; for uniform or worst-case tolerancing one
must instead use bounds such as the $\ell_1$-norm
$\Delta\mathcal{P}_{\max}=\sum_i w_i|J_i|$. Second,
\eqref{eq:genchi2}--\eqref{eq:imhof} assume the quadratic model holds,
i.e.\ that any compensators remain within their linear, unsaturated
stroke (cf.\ Sec.~\ref{sec:demo}E); once a stroke limit binds, the
response is no longer strictly quadratic and the Gaussian tails clip, so
Monte Carlo may again be required to model them. A single
forward-over-reverse AD pass (${\sim}0.5$~s on one GPU\footnote{NVIDIA
1070 Ti.} for the 5-DoF TMA) thus yields the full sensitivity model
deterministically, against the ${\sim}460\times$ cost of an equivalent
Wilks-bounded Monte Carlo (Supplement, Sec.~S3); for $k\!\geq\!9$ DoFs
MC cannot probe the worst-case corner at all.

\subsection{Robust optimisation and Yield Validation}

To demonstrate the practical utility of the proposed robust objective
~\eqref{eq:objective}, we perform a joint optimisation of the
nominal spot size and expected sensitivity on the $5$-DoF alignment
space. We set the weight to $\gamma = 15$ and run the optimisation
starting from the standard nominal design $\bm{p}^*$.
Optimising $J$ improves the manufacturing yield: under Gaussian
tolerances (Fig.~\ref{fig:yield}) the $99.9\%$ spot-size threshold
tightens from $63.11\;\mu\text{m}$ for the standard design to
$55.70\;\mu\text{m}$ for the robust one, an
$11.7\%$ ($1.13\times$) reduction in
worst-case degradation.
On this $5$-DoF problem the gain is dominated by landscape smoothing
rather than desensitization: the penalty guides the single-start
optimiser out of a shallow nominal basin
($\mathcal{P} = 1166\!\to\!247\;\mu\text{m}^2$, RMS
$34.15\!\to\!15.72\;\mu\text{m}$) that a multi-start nominal search
would also escape, while the tolerance trace itself decreases only
marginally ($526.19\!\to\!515.24\;\mu\text{m}^2$). The practical value
of $J$ is thus its differentiability and direct use in gradient-based
loops; genuine desensitization requires robustness slack, which is
absent in directions locked to the optical prescription
(cf.\ Sec.~\ref{sec:demo}E).

\subsection{Scaling and practical challenges}

To demonstrate the scalability of our automatic differentiation
framework, we introduce $18$ high-dimensional surface figure
perturbations: $12$ Fringe Zernike coefficients ($z_{m, j}$ for
  $j=1..4$ representing defocus, astigmatism, and coma errors on all
three mirrors), $3$ conic constants ($k_i$), and $3$ surface
curvatures ($c_i$), yielding a $23$-DoF system. Re-optimising all
$23$ parameters via BFGS yields a near-diffraction-limited baseline
with $\mathcal{P}(\bm{p}^*) = 8.3\;\mu\text{m}^2$ (RMS spot size of
$2.88\;\mu\text{m}$). JAX computes the full, exact $23 \times 23$
Hessian in only $1.8$~s on a single GPU. The tolerance-scaled
eigenvalues span five orders of magnitude, with a sensitivity trace
of $\Tr(\mathbf{S}\mathbf{H}\mathbf{S}) = 1.70 \times 10^{-9}$~m$^2$
($1695.42$~$\mu$m$^2$). Eigendecomposition reveals that the dominant
sensitivity direction (Mode 1: $\lambda_1^{\text{sc}} = 1.61 \times
10^{-9}$~m$^2$) couples the M1 curvature $c_1$ (component $-0.68$)
and inter-mirror spacing $d_1$ ($+0.44$), reflecting their dominant
physical role in system focal length and power folding.
The quadratic model remains predictive at this dimensionality: over
random tolerance-scaled perturbations the Taylor estimate of
~\eqref{eq:taylor} matches the traced degradation to a relative
error below $10^{-3}$ even at full $1\sigma$ amplitude.
We note, however, that this dominant mode is structurally
near-irreducible: because it \emph{is} the system focal length, no
re-optimisation can lower its sensitivity without changing the optical
prescription. Unlike the alignment DoFs of Sec.~\ref{sec:demo}D, this
sensitivity cannot be designed away; it must be met by tighter
manufacturing and assembly tolerances on $c_1$ and $d_1$.
This 23-DoF scaling yields a complete, deterministic sensitivity budget.

In practical systems, active compensation and ray-tracing pipelines
present challenges to standard automatic differentiation. First,
active alignment stages (e.g., focus actuators) are subject to
mechanical travel limits. When a compensator hits a physical bound,
the stationarity condition $\nabla_{\!\bm{c}} L = \bm{0}$ is
violated, breaking the Envelope Theorem's first-order invariance,
so the analytical bounds $\Delta\mathcal{P}_{\max}$ derived above
represent a best-case scenario assuming infinite compensator stroke;
actuator saturation tightens the effective tolerance accordingly and, by
the same token, invalidates the closed-form yield~\eqref{eq:imhof}
(the clipped response is no longer Gaussian), so Monte Carlo is needed
to resolve the saturated tails.
For our TMA, the natural compensator is detector refocus along the
post-M3 axis; a $1{,}000$-sample tolerance Monte Carlo shows the
optimal refocus demand stays below $0.4\;$mm (99.9th percentile),
an order of magnitude inside the stroke of a standard focus stage,
so the compensator remains interior and the first-order invariance
holds here. Saturation would bind only under coarser tolerances or
shorter-stroke actuators.
In these constrained regimes, one must project the Hessian onto the
active constraint boundary (e.g., via active-set projection) to
isolate the constrained Schur complement.

In both demonstrations, all $1{,}800$ traced rays remain within
each mirror's physical aperture throughout the optimisation; no
binary clipping is applied. The mirror radii ($R_1$–$R_3$) enter
the kernel solely as Zernike normalization constants.
When footprint spillover is present, differentiability can be restored
via sigmoidal soft-clipping or randomized smoothing.

\section{Conclusion}

We have demonstrated that the tolerance Hessian provides a rigorous second-order
characterization of sensitivity to manufacturing tolerances. The analysis of
perturbation modes yields qualitative insight into the system response, revealing
coupling mechanisms that are not captured by classical first-order sensitivity
matrices. Its primary utility lies in defining a deterministic sensitivity budget: the
trace of the tolerance-scaled Hessian, together with its eigenmode decomposition,
quantifies expected performance degradation and identifies the dominant directions of
sensitivity. We further show that a single closed-form evaluation replaces costly and
inaccurate Monte Carlo sampling, enabling efficient and targeted allocation of
manufacturing tolerances. When used as an optimization penalty, the proposed metric
promotes robust designs in regimes where sufficient slack is available; otherwise, it
directly identifies the critical tolerances that must be tightened to meet performance
specifications.


\begin{backmatter}
  \bmsection{Funding}
  Thales Alenia Space.

  \bmsection{Acknowledgment}
  The authors would like to thank D. Serre, N. Tétaz, V. Laborde and A. Bellaiche at
  Thales Alenia Space for fruitful discussions and feedback about this work.

  \bmsection{Disclosures}
  The authors declare no conflicts of interest.

  \bmsection{Data Availability Statement}
  Data underlying the results presented in this paper are not
  publicly available at this time but may be obtained from the
  authors upon reasonable request.

  \bmsection{Supplemental document}
  See Supplement~1 for supporting content, including the full TMA
  prescription required to reproduce the numerical experiments, the
  second-order (Schur-complement) compensator derivation, and the
  Monte Carlo cost analysis.
\end{backmatter}

\bibliography{sample}
\bibliographyfullrefs{sample}

\end{document}